# Efficient Discrete Feature Encoding for Variational Quantum Classifier

**HIROSHI YANO[1]** , **YUDAI SUZUKI[2]**, **KOHEI M. ITOH[1,3]**,
**RUDY RAYMOND[3,4]** , **AND NAOKI YAMAMOTO[1,3]**

[1] Department of Applied Physics and Physico-Informatics, Keio University, Yokohama 223-8522, Japan
[2] Department of Mechanical Engineering, Keio University, Yokohama 223-8522, Japan
[3] Quantum Computing Center, Keio University, Yokohama 223-8522, Japan
[4] IBM Research—Tokyo, Tokyo 103-8510, Japan

Corresponding authors: Hiroshi Yano; Rudy Raymond (hiroshi.yano.gongon@keio.jp; rudyhar@jp.ibm.com).

This work was supported by MEXT Quantum Leap Flagship Program under Grant JPMXS0118067285 and Grant JPMXS0120319794.

**ABSTRACT** Recent days have witnessed significant interests in applying quantum-enhanced techniques for solving a variety of machine learning tasks. Variational methods that use quantum resources of imperfect quantum devices with the help of classical computing techniques are popular for supervised learning. Variational quantum classification (VQC) is one of such methods with possible quantum advantage in using quantum-enhanced features that are hard to compute by classical methods. Its performance depends on the mapping of classical features into a quantum-enhanced feature space. Although there have been many quantum-mapping functions proposed so far, there is little discussion on efficient mapping of discrete features, such as age group, zip code, and others, which are often significant for classifying datasets of interest. We first introduce the use of quantum random-access coding (QRAC) to map such discrete features efficiently into limited number of qubits for VQC. In numerical simulations, we present a range of encoding strategies and demonstrate their limitations and capabilities. We experimentally show that QRAC can help speeding up the training of VQC by reducing its parameters via saving on the number of qubits for the mapping. We confirm the effectiveness of the QRAC in VQC by experimenting on classification of real-world datasets with both simulators and real quantum devices.

## I. INTRODUCTION

Progress in quantum computing hardware has accelerated research in using quantum computing for machine learning tasks [1]–[8], such as classification, regression, and anomaly detection. Much of the quantum-enhanced machine learning techniques, especially for classification, are derived from the variational methods [9], [10], which are also popular in optimization [11]–[13]; those methods are believed capable of utilizing noisy quantum devices [14] and classical computing devices to obtain quantum advantage. The quantum-enhanced techniques can be classified into two: a direct approach using variational quantum classifier (VQC) and an indirect approach using quantum kernel estimation [5], [15]. In brief, the two methods are for mapping real-valued features into quantum-enhanced feature space whose mapping is believed to be difficult to compute by any classical computational devices under some computational complexity arguments [5].

However, the majority of mapping techniques deal with real-valued feature vectors, whereas those for binary and discrete features are lacking. On the other hand, although quantum feature space is often associated with analog features of quantum bits that can store continuous values, the power of quantum bits is limited. For example, if the information is to be recovered with certainty, Holevo bound restricts that $n$ qubits can only store up to $n$ bits of information, and nothing more [16]. The restriction even hold in the probabilistic setting due to Nayak bound [17] that limits the amount of information retrievable from $m$ qubits to recover any one out of $n$ bits (for $n \geq m$). Namely, one still needs $n$ qubits in order to recover any one out of $n$ bits with certainty. A linear saving is possible if constant errors are allowed in retrieving any one of the bits, but no more advantage is offered by quantum-enhanced coding in this case.

Nevertheless, if the error probability is allowed to grow with the number of qubits while limited there are advantages offered from encoding bits into less number of qubits, which is termed as quantum random-access coding (QRAC) [18], [19]. QRAC is an encoding of bitstring of length $n$ into $m$ qubits so that any one out of $n$ bits can be recovered with probability at least $p > 1/2$. Such QRACs are often denoted as $(n, m, p)$-QRACs. They have been used for demonstrating many possible quantum advantages in quantum communication protocols and others [20]–[22]. There are also research on how to construct such QRACs for specific number of qubits [23], [24], in addition to asymptotic values of $n$ [18]. At the heart of QRAC is mapping bitstrings into quantum states so that the distance between any two quantum states is proportional to their bitstrings' Hamming distance. This makes QRACs good candidates for mapping discrete features.

Classification models in real-world datasets often also depend on binary features, such as yes–no answers to questions, in addition to (discrete) categorical features that are naturally represented with bitstrings, such as zip code, age, and ethnic group. Such discrete features have to be encoded into continuous features before they can be used effectively in machine learning models that rely on continuity of their inputs. There have been many proposed encodings, with *one-hot* encoding as one of the most popular encodings, for such purposes [25]. It is known that the encodings can heavily impact the performance of the learning models (see Section II for more details). Mapping such binary and discrete features into quantum-enhanced feature space is not trivial and to our knowledge there is not much discussion for them despite their significance in classification and other machine learning tasks, and the fact that VQC assumes the continuity of its input features.

In this work, we contribute in three major aspects of mapping discrete features for VQCs with focus on supervised learning. First, we propose QRACs for mapping binary and categorical features into quantum-enhanced feature space in the mechanism of VQCs. There are many classical strategies developed in machine learning to map discrete features into continuous ones, but QRACs can be used to map them more efficiently with less number of qubits in addition being a mapping that is uniquely quantum. Second, we provide some technical analysis of QRACs for classification tasks, which we hope can provide new insights in mapping discrete features into quantum-enhanced feature space. Finally, we show the effectiveness of utilizing QRACs in VQCs for practical machine learning problems in applications where discrete features play important roles. We confirm with experiments on both simulator and quantum devices that the resulting VQCs can achieve better classification results in less training time than traditional VQCs. This is due to using less number of qubits in the quantum mapping and, hence, less number of parameters required for tuning during the training phase of the VQCs with QRACs. Our contributions shed new light on practical applications of QRACs despite the theoretical limitation of the advantage of QRACs against their classical counterparts, as shown recently by Doriguello and Montanaro [26] in the context of *functional* QRACs.

The rest of this article is organized in the following order. We list related work in Section II with emphasis on feature mapping techniques. We then review the VQC in Section III-B and QRAC in Section IV. In particular, our proposed method of employing QRACs into VQCs is detailed in Section IV-B, and in Section IV-C, we mention the relation to the functional QRACs [26], which can be viewed as binary classification with binary inputs. In Section V, we present numerical demonstration of encoding binary features with QRACs in several forms and provide arguments on the possibilities and limitations of QRACs for supervised learning. We present experimental comparisons of the proposed method against standard VQC on some benchmark datasets in Section VI. Finally, Section VII concludes this article.

## II. RELATED WORK

Structured datasets with discrete features are omnipresent. The discrete features are referred to categorical or qualitative data. They are important features that heavily impact the performance of prediction models. While discrete features can be used naturally in some learning models, such as decision trees [27], in the most popular neural-network model [28], they must be first transformed into continuous features. In fact, it is observed that even though neural-network models are prominent for dealing with unstructured datasets, they are less, so for structured datasets with categorical features as observed in [29], that tree-based models are popular choices of many winning teams in machine learning competition.

Techniques to use categorical features in neural network models whose inputs are of continuous nature are important. The continuity of features is necessary to guarantee convergence in the training phase and stability of output under slight changes of inputs in the prediction phase. Because straightforward use of integers replacing categorical features does not work well in neural network models, there are quite a variety of classical techniques to map categorical values to numerical values [25]. They are also known under different names: (entity) embeddings [29], (dense) encodings [30], and (distributed) representations [31].

Based on the degree of the complexity, the mapping techniques can be classified into three categories: determined, algorithmic, and automated [25]. Determined techniques are the simplest and include the most popular one-hot encoding, ordinal coding, hash encoding, etc. Techniques in this category fixed the encoding of the categorical values based on some simple rules or lookup tables. For example, the one-hot encoding represents $d$ distinct categorical values with bitstrings of length $d$ with single 1 and all the others 0, such as 100, 010, and 001 when $d = 3$. They are widely used and their implementation is available in popular machine learning library, such as scikit-learn [32].

The algorithmic techniques use more advanced preprocessing steps, which often involve other machine learning models. Their output vectors are often heuristics and tailored to specific application domains, such as latent Dirichlet allocation [33], which is popular in natural language processing. The automated techniques are the most complicated and resource intensive. They can find the vector representation of categorical features tailored to the distribution of inputs. They involve neural networks to generate representations in end-to-end manner. Word2vec [34], which is one of the most celebrated distributed representation of (discrete) words, is an example of automated techniques. Both algorithmic and automated techniques often utilize determined techniques, such as one-hot encoding, as their inputs.

Meanwhile, quantum-enhanced machine learning techniques are very similar to neural network models and, thus, not surprisingly many of their frameworks [35], [36] are inspired by classical neural-network framework [37]. Like their classical counterparts, most known quantum-enhanced machine learning techniques assume continuous features, i.e., real-valued vectors, whereas techniques of quantum machine learning [1] are mostly from quantum basic linear algebra subprograms, such as the prominent HHL algorithm [38]. Those quantum-enhanced subprograms heavily rely on appropriate representation of input datasets.

The embedding of classical data into the vast Hilbert space of quantum system is a central topic in utilizing kernel tricks with quantum-enhanced support vector machine (SVM) on near-term quantum devices [5], [8], [15], [39]. Nevertheless, to our knowledge, quantum methods to deal with categorical features are lacking. This is perhaps because one can utilize aforementioned classical techniques to encode them before being used in the quantum subprograms. Indeed, there are recent proposed methods [36], [40] that combine classical neural networks with quantum models for classification and other machine learning tasks.

Schuld and Killoran [15] were among the first to consider encoding inputs into quantum state as feature maps. Their consideration led to two ways of building quantum classifiers for supervised learning: an implicit approach through kernel functions evaluated by quantum devices, and an explicit approach through learning linear decision boundary in the quantum feature space. Both approaches nonlinearly map the data $x$ to a pure quantum state with $\Phi : x \rightarrow \Phi(x)\rangle$. Several input encodings as feature maps, such as basis encoding, amplitude encoding, copies of quantum states, and product encoding, are proposed in [15]. The basis encoding, which trivially maps bitstrings to their corresponding computational basis, can be used for encoding discrete features but it requires $n$ qubits for mapping $n$ bits. The amplitude encoding [8], which maps normalized real-valued vectors into superposed quantum states with probability amplitudes proportional to the elements of the vectors, can encode $n$ bits with $\log n$ qubits and perhaps the closest to our proposed encoding. However, our proposed QRACs requires only $\log(n)/2$ qubits (half of that in amplitude encoding). We note that we would be able to use the amplitude encoding for the bitstring data, but, in general, the amplitude encoding requires exponential number of gates to generate the target quantum state [41]–[44]: in this general sense, the cost of encoding classical real-valued data into the amplitude of a quantum state may diminish the possible advantage of quantum computation (as often discussed in this research field). The copy encoding and product encoding correspond to, respectively, the well-known polynomial and cosine kernels. However, they are mostly for real-valued features, such as other similar nonlinear feature maps, i.e., *squeezing* in continuous quantum systems [15], *density-operator* encoding [39], and *quantum metric learning* [45]. A recent overview of data encodings and quantum kernels can be found in [46]. It is not trivial how to use them to encode discrete features.

Our main tool to encode a $n$-bit string into $\log(n)/2$ qubits is the QRAC. QRAC is one of examples in which quantum schemes are better than their classical counterparts. QRACs can encode bits with the number of qubits half of the bits used in classical random-access codings (RACs). The halving advantage offered by QRACs is similar to superdense coding [47] and quantum teleportation [48]. Originally formulated in the communication setting, QRACs have been extensively used in the theory of quantum computations, such as the limit of quantum finite automata [19] and quantum state learning [49]. QRACs are applied in quantum communication complexity [50] and are used in elaborate coding schemes, such as network coding [21] and locally decodable codes [51]. QRACs have also been applied in quantum nonlocality and contextuality [52], cryptography [53], and random number generation [54]. Some QRACs also offer cryptographic properties known as *parity obliviousness* [55], which can play important role in cryptography and private information retrieval.

There are variants of QRACs using shared entanglement and classical randomness [56], which enable encoding any number of bits into a single qubit, and using $d > 2$-level quantum systems (e.g., qutrits) [23]. Experimental realization of QRACs on few quantum resources has been shown [52]. More variants of QRACs can also be found in the recent study [26], which confirms the theoretical advantage of QRACs is asymptotically the same as their classical counterparts.

We focus on QRACs with few qubits because larger QRACs used in theoretical studies are essentially concatenation of smaller QRACs. Such QRACs were first given by Wiesner [57] and popularized by Ambainis *et al.* [19], which showed explicit constructions of (2,1,0.85)-QRAC. The QRAC for encoding 3 b of information into one qubit was attributed to Chuang whose construction of (3,1,0.78)-QRAC was shown in [58]. QRACs with two qubits were shown in [23] and [24]. For more qubits, the generic construction of $(O(m), m, > 1/2)$-QRACs was first introduced by Ambainis *et al.* [19], but it was essentially RACs. The first generic construction of $(n, m, p)$-QRACs for any $n < 2^{2m}$ was shown in [21]. There are some well-known limitations of QRACs. For example, any $(n, m, p)$-QRACs must satisfy *Nayak bound* [17]: $m \geq (1 - H(p))n$, where $H(\cdot)$ is the entropy function. Moreover, $n$ cannot exceed $2^{2m} - 1$, as shown in [58]. Thus, one qubit can only encode at most 3 b, and two qubits at most 15 b, etc.

## III. VARIATIONAL QUANTUM CLASSIFIER

In the following, we briefly describe the model of VQC particularly for binary classification problems, where two labels are obtained by a two-valued quantum measurement. Note that we can extend the classifier for multiclass classification, as suggested in [5] and [59]. For example in [59], the $n$-class classification is achieved by computing the overlap of the final state with $n$ maximally orthogonal target states, each one corresponding to a label. This can be regarded as the decoding process of QRAC since the target state is determined with multiple measurements.

### A. CLASSICAL SVM

Assume that we are given the training dataset $S = \{(x_1, y_1), (x_2, y_2), \ldots, (x_{m_S}, y_{m_S})\}$, where $x_i \in \mathbb{R}^d$ and $y_i \in \{-1, 1\}$. The goal of learning a binary classifier from $S$ is to construct a function $f(x)$ so that $f(x_i)y_i > 0 \quad \forall i$. The simplest form of such function is a linear classifier $f(x) = w^T x + b$, where $w \in \mathbb{R}^d$ and $b \in \mathbb{R}$. $S$ is called linearly separable if there is a $(w, b) \in \mathbb{R}^{d+1}$ satisfying $f(x_i)y_i > 0$ for every $i$. In general, there can be many classifiers that separate the training dataset exactly; a reasonable classifier is the one that maximizes the distance with respect to the closest points of the two classes. It is known that finding such classifier is reduced to solving the quadratic optimization problem known as hard-SVM [60].

The hard-SVM can be relaxed to produce a classifier that predicts the training dataset almost correctly so that $f(x_i)y_i \geq 1 - \epsilon_i$ for all $i$. It is referred to as the soft-SVM [60]. The slack variables $\{\epsilon_i \geq 0\}$ determine the quality of the classifier; the closer they are to zero the better the classifier. For this purpose, we can embed the data $\{x_j\}$ into a larger space by finding a map $x : \Phi(x) \in \mathbb{R}^n$ for $n > d$. The classifier $f(x)$ is now defined as $f(x) = w^T \Phi(x) + b$. When $\Phi(x)$ is an embedding of data nonlinearly to quantum state $|\Phi(x)\rangle$, then we can use quantum-enhanced feature space for the classifier: this is the core idea of VQC.

### B. QUANTUM-ENHANCED VARIATIONAL CLASSIFIER

VQC relies on techniques for finding the best hyperplane $(w, b)$ that linearly separates the embedded data. First, the data $x \in \mathbb{R}^d$ are mapped to a (pure) quantum state by the feature map circuit $U_{\Phi(x)}$ that realizes $\Phi(x)$. This means that, conditioned on the data $x$, we apply the circuit $U_{\Phi(x)}$ to the $n$-qubit all-zero state $|0_n\rangle$ to obtain the quantum state $|\Phi(x)\rangle$. A short-depth quantum circuit $W(\theta)$ is then applied to the quantum state, where $\theta$ is the vector composed of parameters that will be learned from the training data. Finding the circuit $W(\theta)$ is akin to finding the separating hyperplane $(w, b)$ in the soft-SVM, with the promise of quantum advantage that stems from the difficulty for classical procedures to realize the feature map $\Phi(x)$.

The binary decision is made by measuring the final quantum state in the computational basis to obtain $z \in \{0, 1\}^n$ and linearly combining the measurement results; this is equivalent to measuring $g = \sum_{z \in \{0,1\}^n} g(z)|z\rangle\langle z|$, where $g(\cdot) \in \{-1, 1\}$. The probability of measuring $z$ is given as

$$|\langle z| W(\theta) |\Phi(x)\rangle|^2 = \langle\Phi(x)| W^\dagger(\theta) |z\rangle\langle z| W(\theta) |\Phi(x)\rangle.$$

The function $f(x)$ is then given by the mean of $g$ with the bias $b$

$$f(x) = \langle\Phi(x)| W^\dagger(\theta) g W(\theta) |\Phi(x)\rangle + b. \tag{1}$$

The predicted label is then given by the sign of $f(x)$. The hyperplane $(w, b)$ is now parameterized by $\theta$. The $i$th element of $w(\theta)$ is $w_i(\theta) = tr(W^\dagger(\theta) g W(\theta) P_i)$, where $P_i$ is a diagonal matrix whose elements are all zeros except the $(i, i)$ element, which is one. Also the $i$th element of $\Phi(x)$ is $\Phi_i(x) = \langle\Phi(x)| P_i |\Phi(x)\rangle$.

Learning the best $\theta$ can be obtained by minimizing the cost function formulated as empirical risk $R(\theta)$ or binary cross entropy $H(\theta)$ with regards to the training data $S$. These cost functions to be minimized are

$$R(\theta) = \frac{1}{|S|} \sum_{i \in [m_S]} |f(x_i) - y_i| \tag{2}$$

$$H(\theta) = -\frac{1}{|S|} \sum_{i \in [m_S]} \left( \frac{1 - y_i}{2} \log(p_i^{(-1)}) + \frac{1 + y_i}{2} \log(p_i^{(1)}) \right) \tag{3}$$

$$\text{where}: p_i^y = \sum_{z \in \{0,1\}^n | g(z) = y} |\langle z| W(\theta) |\Phi(x)\rangle|^2.$$

The empirical risk can be approximated by a continuous function using sigmoid function, as detailed in [5]. This enables applying a variational method, such as COBYLA or SPSA, for tuning $\theta$ to minimize the approximated cost functions.

The binary classification with VQC now follows from the training of the classifier to learn the best $\theta^*$, which minimizes the cost function and gives $(w(\theta^*), b^*)$. The classification for unseen data $x$ is then performed according to the classifier function $f(x)$ with $(w(\theta^*), b^*)$. Both training and classification need to be repeated for multiple times (or shots) due to the probabilistic nature of quantum computation. The former may need significant number of shots proportional to the size of $S$ but it can be performed in batch offline. On the other hand, the latter needs much less number of shots, and may be performed online (or near real time).

We have described the procedure for designing a classifier in a general form. In this article, we particularly limit the data $x$ to a vector of discrete variables (or the mixture of continuous and discrete variables). Hence, our problem is how to create an efficient feature map circuit $U_{\Phi(x)}$ in such a special case. We will discuss our method in Section IV.

## C. NONLINEAR EMBEDDING

There are many classical methods for nonlinear embedding of data $x$ : $\Phi(x) \in \mathbb{R}^n$ for $n > d$, such as the polynomial kernel, which is also popular for natural language processing [61]. In this case, the two-dimensional data $(x_1, x_2)$ is embedded into a three-dimensional one $(z_1, z_2, z_3)$ such that $z_1 = x_1^2$, $z_2 = \sqrt{2}x_1 x_2$, and $z_3 = x_2^2$. On the other hand, in the quantum-enhanced SVM, the embedding of data to the $n$-qubit feature space is performed by applying the unitary $\mathcal{U}_{\Phi(x)} = U_{\Phi(x)} H^{\otimes n} U_{\Phi(x)} H^{\otimes n}$, where $H$ is the Hadamard gate, and $U_{\Phi(x)}$ denotes a diagonal gate in the Pauli-$Z$ basis as follows:

$$U_{\Phi(x)} = \exp\left( i \sum_{S \subseteq [n]} \phi_S(x) \prod_{k \in S} Z_k \right) \qquad (4)$$

where the coefficients $\phi_S(x) \in \mathbb{R}$ are fixed to encode the data $x$. For example, for $n = d = 2$ qubits, $\phi_i(x) = x_i$ and $\phi_{1,2}(x) = (\pi - x_1)(\pi - x_2)$ were used in [5]. The classification performance greatly depends on these functions [62], [63]. In general, $U_{\Phi(x)}$ can be any diagonal unitary that is efficiently realizable with short-depth quantum circuits. In total, one needs at least $n \geq d$ qubits to construct such quantum-enhanced feature map, i.e., the number of qubits is at least the dimension of the feature vector of the datasets.

# IV. CLASSIFICATION ON DISCRETE FEATURES

In this section, we describe the concept of QRAC and how it can be used to map binary features with less number of qubits in VQC. When the inputs are binaries, the classification can be regarded as evaluating Boolean functions, which coincides with the functional QRAC (or $f$-QRAC) [26]. We discuss the relation of our proposal with $f$-QRAC, which will be applied to VQC in the next section.

## A. DEFINITION OF QRAC

QRACs are coding schemes that allow us to encode $n$ bits into $m$ qubits, for $n > m$, so that any one of the bits can be extracted with success probability at least $p > 1/2$. More precisely, the $(n, m, p)$-QRAC is a function that maps a $n$-bit string $b \in \{0, 1\}^n$ to a $m$-qubit state $\rho_b$ such that for every $i \in \{1, 2, \dots, n\}$, there exists a positive operator-valued measure (POVM) $E^i = \{E_0^i, E_1^i\}$ satisfying $\mathrm{Tr}(E_{b_i}^i \rho_b) \geq p$ for all $b \in \{0, 1\}^n$; that is, the probability of retrieving any $i$th bit $b_i$ is at least $p$.

Similarly, one can define a classical encoding function under the same setting as the QRAC, namely the $(n, m, p)$-RAC, to use $m$ bits to encode $n$ bits of information so that each of the bits can be recovered with probability at least $p$. QRAC is an example where quantum systems are better than their classical counterparts. Namely, in the QRACs, the number of required qubits is half of the bits used in the classical RACs; that is, $(n, m, p)$-QRACs exist for any $n < 2^{2m}$, whereas their classical counterparts $(n, m, p)$-RACs exist only for $n < 2^m$.

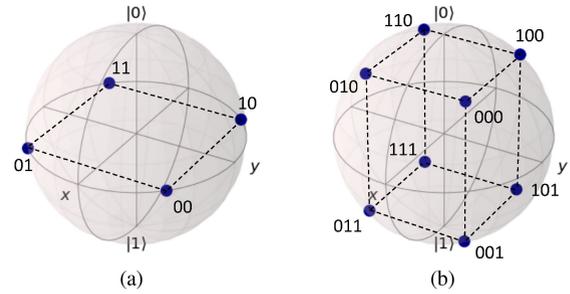

**FIG. 1.** Bloch sphere representation of $(2, 1, 0.85)$-QRAC and $(3, 1, 0.78)$-QRAC. Each bitstring is mapped on the surface of the sphere. The distance between two quantum states is proportional to the Hamming distance of their corresponding bitstrings. (a) $(2, 1, 0.85)$-QRAC. (b) $(3, 1, 0.78)$-QRAC.

The optimal QRACs for encoding up to 3 b into one qubit are known: $(2, 1, 0.85)$-QRAC and $(3, 1, 0.78)$-QRAC. The Bloch sphere representation of these QRACs is shown in Fig. 1. It is also known how to construct QRACs for encoding up to 15 b into two qubits, such as the $(3, 2, 0.90)$-QRAC [24]. However, the general way to construct QRACs is not known. In what follows, we often use $(n, m)$-QRAC to refer to $(n, m, p)$-QRAC, when the meaning of $p$ is clear from the context.

## B. MAPPING DISCRETE FEATURES WITH QRAC

Let us consider the dataset $\{(x_i, y_i)\}_{i=1,\dots,m_S}$ with variable $x_i$ and label $y_i$. The variable $x_i$ is further partitioned into variables that represent the discrete and continuous parts, each represented as $x_i^{(b)}$ and $x_i^{(r)}$. The discrete part $x_i^{(b)}$ is obtained by encoding the categorical feature into the bitstring via, e.g., the one-hot encoding.

We propose to use QRACs for directly encoding the bitstring $x_i^{(b)}$ to a quantum state. For any bitstring of length $n$, we can use at most $m = \lceil \log(n)/2 \rceil$ qubits, provided we have the constructions of $(n, m, p)$-QRACs. Moreover, the quantum states obtained from the QRACs used to represent any two different bitstrings $x_i^{(b)}$ and $x_j^{(b)}$, say $\rho_{x_i^{(b)}}$ and $\rho_{x_j^{(b)}}$, preserve some properties of the bitstrings. For example, for $p \geq 1/2 + \epsilon$, it is guaranteed that the trace distance between those quantum states satisfy $D(\rho_{x_i^{(b)}}, \rho_{x_j^{(b)}}) \geq 2\epsilon$, because QRACs guarantee that there is a measurement (at the indices where the two bitstrings differ) distinguishing the two quantum states with the margin $\epsilon$ from $1/2$. Moreover, the more number of bits differ, the more measurements are available to distinguish the quantum states.

However, QRACs impose that each one out of $n$ bits is recoverable with probability better than random guess. This condition requires us to prepare $n$ different types of measurement for the decoding. On the other hand, for the binary classification task, not all measurements are necessary and it suffices to find a single measurement whose results are then used in conjunction with the measurement result of the continuous variables to classify the data. In the VQC framework, this measurement operator is realized by the parameterized circuit $\mathbf{W}(\theta)$ followed by the computational basis measurement. In the following, we refer to the VQC with QRAC as simply VQC+QRAC, as shown in Fig. 2. Also, we show the schematic of VQC without QRAC in Fig. 3, which will be compared to VQC+QRAC in the next section.

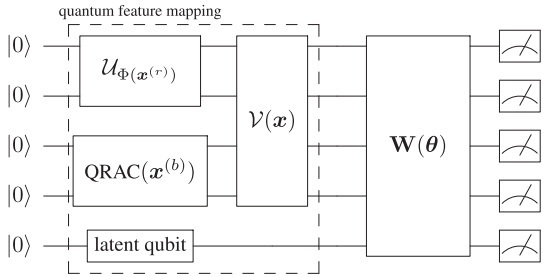

**FIG. 2.** Quantum circuit for VQC+QRAC for encoding discrete features. The additional gate $\mathcal{V}(x)$ may be included for entangling the quantum states for continuous variables and discrete variables. Also, latent qubits may be included to extend the dimension of the Hilbert space.

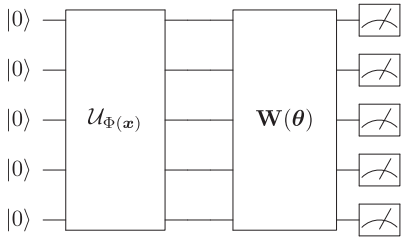

**FIG. 3.** Quantum circuit for VQC without QRAC, which consists of fixed quantum feature mapping $\mathcal{U}_{\Phi(x)}$ and the separator $\mathbf{W}(\theta)$ trained with the variational method.

### C. RELATION TO FUNCTIONAL QRAC

Doriguello and Montanaro [26] extended the idea of QRAC to recovering the value of a given Boolean function $f$ on any sequence of $k$ bits from the input bitstring $x$. They call this scheme the $f$-QRAC. Here, we briefly review the $f$-QRAC and explain how it relates to our work.

Instead of recovering an input bit, the $f$-QRAC asks for recovering the value of a Boolean function $f$, which depends on a fixed size of the input bits, with probability $p > 1/2$. The $f$-QRAC is a generalization of recovering multiple bits by the QRACs instead of a single bit, which has been considered by Ben-Aroya *et al.* [64] for parity functions. In [26], it is shown that the success probability for $f$-QRAC is characterized by the resiliency of the Boolean function to the changes of its input bits (or noise stability), which then can be used to compare them with the classical counterparts. It turns out that the theoretical quantum advantage is only a factor of multiplicative constant against the classical.

The multiplicative constant advantage is readily available by the concatenation of QRACs with one qubit, as similarly done by Pawłowski *et al.* [65], [66]. The key idea of constructing $f$-QRAC is to randomly divide $n$ bits of $x$ into $m$ sets and encode them with $m$ sets of $(n/m, 1)$-QRAC; that is, each set is encoded into a single qubit. These $m$ copies are used to amplify the success probability for extracting a particular bit. In this scheme, all the $k$ bits of the substring of $x$ will be encoded into different sets with high probability, which allows to decode the significant substring of $x$ with probability $p > 1/2$ in the end, and hence to compute $f(x)$ correctly with the same probability. This recent result of $f$-QRAC provides us a new insight to implement VQC+QRAC for classification in at least two ways.

First, it gives a theoretical foundation to deal with longer input bits by concatenating smaller QRACs to obtain modest quantum advantage. Thus, for mapping discrete features in VQC+QRAC, any $(n, m, p)$-QRACs can be used and we can, of course, develop them with $m$ sets of $(n/m, 1)$-QRAC by following $f$-QRAC. If all the $k$ bits of the substring of $x$ are independently encoded into different sets, the result of $f$-QRAC that the value of $f$ is correctly evaluated with probability $p > 1/2$ suggests that for the VQC, there exists a measurement operator $\mathbf{W}^{\dagger}(\theta) g \mathbf{W}(\theta)$ in (1) that perfectly classifies the data points in the $m$-qubit Hilbert space. We will confirm this numerically in Section V-C. Note that in the $f$-QRAC, the value of the Boolean function is evaluated classically after decoding the determining bits, whereas in the VQC, the value of the Boolean function is directly evaluated from the quantum circuits with a measurement operator. Quantum circuits for $f$-QRAC and VQC+$f$-QRAC (will be called VQC+QRAC for simplicity) are depicted in Fig. 4.

Second, it attracts an investigation on the practicality of concatenating smaller QRACs to encode longer bits. Although in theory, the asymptotical advantage of concatenating QRACs on one qubit is the same as those on two qubits, it is important to see how we can get a bigger constant factor of qubit savings by employing two-qubit QRACs. Not unsurprisingly, we find that for some Boolean functions, in practice, it may be better to use two-qubit QRACs instead of concatenation of one-qubit QRACs, as detailed in Section V-C. This hints the importance of concrete constructions of QRACs, which has received less spotlight in the theoretical studies.

## V. POTENTIAL OF VQC+QRAC

This section demonstrates VQC+QRAC in various forms and discusses its capabilities, with the use of classical simulations of quantum computation via Qiskit [35]. We slightly modified VQC implemented in Qiskit for mapping binary features with QRAC to obtain VQC+QRAC. In this simulation and the following experiments in Section VI, we trained both VQC and VQC+QRAC models to minimize the cross entropy (3) instead of the empirical risk (2). The bias is $b = 0$ in (1). The parameterized circuits $\mathbf{W}(\theta)$ of both models are built from the so-called RyRz ansatz, which is composed of layers of fully connected $CZ$ gates and single-qubit ($y$ and $z$) rotation gates. Note that this ansatz is used by Havlicek *et al.* [5] as well.

### A. VQC WITH ONE-QUBIT QRAC

First, we show the performance of VQC+QRAC in the simplest forms, with (2,1)-QRAC or (3,1)-QRAC. Recall that at the heart of VQC+QRAC is mapping discrete features into the quantum state composed of smaller number of qubits than the classical case. This can be disadvantageous, because the mapping can result in the arrangement of points that are not separable by any hyperplane in the small Hilbert space. Here, we will investigate this limitation and discuss how to handle this problem.

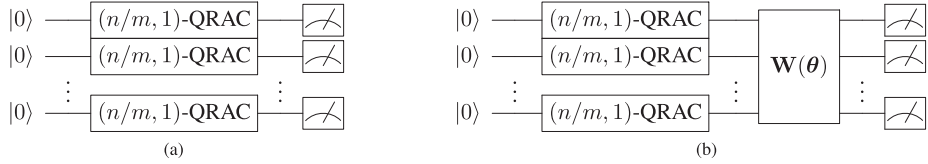

**FIG. 4.** Quantum circuits for *f*-QRAC and VQC+QRAC. (*n*, *m*)-QRAC can be realized with *m* sets of (*n/m*, 1)-QRAC. (a) Measurement outcomes are used to construct the significant substring with probability $p > 1/2$, thereby the value of the Boolean function is classically evaluated. (b) Idea of *f*-QRAC is based on the fact that there exists a measurement basis to decode the significant bit with high probability for each (*n/m*, 1)-QRAC. Such measurement bases can be found with a parameterized circuit $\mathbf{W}(\theta)$ in the VQC+QRAC setting. This leads to the existence of the measurement operator in VQC that perfectly classifies the data points in the *m*-qubit Hilbert space when a bitstring is encoded with *f*-QRAC.

For example, the (2,1)-QRAC maps the 2-b features into four points on a two-dimensional plane (as in [18] and [58]). We can show that there are arrangements of four points, which cannot be separated by any hyperplane. This can be explained using Vapnik–Chervonenkis (VC) dimension [67]. The VC dimension represents the learning capacity of a class of functions and is defined as the maximum number of training points that can be shattered by a class of functions. Because the VC dimension of linear separators in $\mathbb{R}^2$ is 3, there exist cases where the four points cannot be shattered by any linear classification model in the two-dimensional plane. The same holds for the case of the (3,1)-QRAC that maps 3-b features into eight points in a three-dimensional space. Because the VC dimension of linear separators in $\mathbb{R}^3$ is 4, there exist cases where a mapping results in failed classification. The reference for the aforementioned discussion can be found in [68].

The aforementioned limitation can be clearly seen in Fig. 5, showing the problem of classifying blue and orange data points (each representing 0 and 1 classification label), which are mapped on the surface of Bloch sphere using QRAC. Note that the datasets in Fig. 5 are all possible configurations for (2, 1)-QRAC and (3, 1)-QRAC, considering the geometrical symmetry of the sphere. We name each pattern as type *a-b-c*, which represents the pattern number *c* of (*a*, 1)-QRAC with *b* blue points and $2^a - b$ orange points. We can see from the figure that one-qubit VQC-QRAC cannot separate data points of some patterns, i.e., type 2-2-2 for (2, 1)-QRAC and type 3-2-2, 3-2-3, 3-3-2, 3-3-3, 3-4-3, 3-4-4, and 3-4-5 for (3, 1)-QRAC; this fact still holds even when optimizing the bias parameter *b*, which is equivalent to moving the hyperplane in a direction perpendicular to it. This impossibility is in fact easily predicted, because there is no plane (i.e., measurement operator) that can perfectly separate data points in those patterns, as seen in Fig. 5.

To overcome this limitation of VQC+QRAC, we need to increase the dimension of the feature space. One solution to the inseparability in type 2-2-2 is to use (3, 1)-QRAC instead, which means mapping the features into higher dimension. There are at least three means to add dimensionality. First, by adding latent qubits, as suggested in [45], which are qubits initialized to some fixed quantum states, as shown in Fig. 2. Second, by using multiple copies of QRACs for encoding the same discrete features, as will be seen in Section VI-B1, which is similar to using copies encoding suggested in [39]. Third, we may use higher dimensional QRACs, such as (3,2,0.91)-QRAC shown in [24], for mapping eight points into two-qubit Hilbert space. We ran experiments comparing the effectiveness of these three methods and obtained the train loss curves for type 3-4-5, as shown in Fig. 6. We confirmed that all methods overcome the limitation of the QRAC; in particular, the use of higher dimensional QRACs seems to be the most effective. In fact, two-qubit VQC+QRAC with the copies of (3,1)-QRAC and (3,2)-QRAC can completely separate the data points of type 3-4-5. We also confirmed that by adding dimension to the inseparable types of Fig. 5, there are many cases where all points can then be shattered.

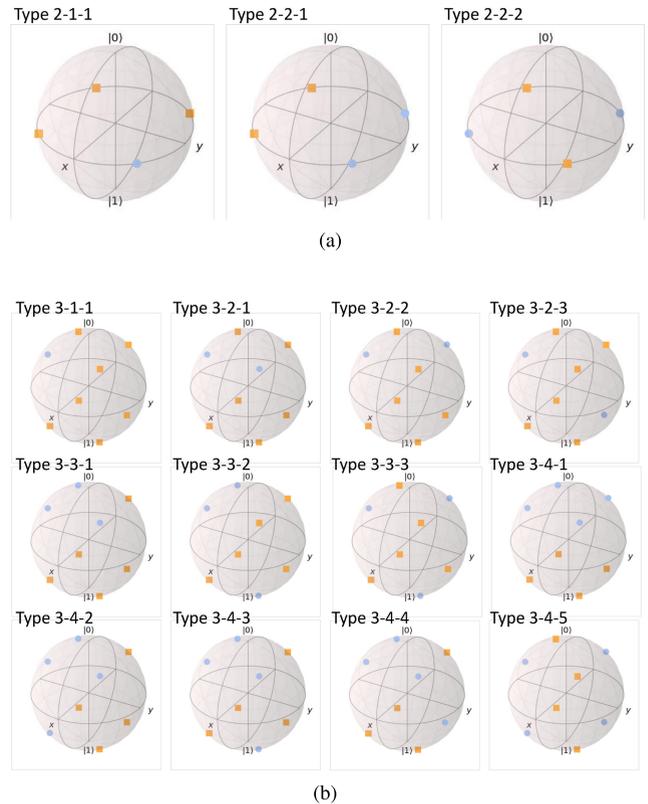

**FIG. 5.** All possible configurations of (2, 1)-QRAC and (3, 1)-QRAC. Type *a-b-c* represents the pattern number *c* of (*a*, 1)-QRAC with *b* blue points and $2^a - b$ orange points. (a) Case of (2, 1)-QRAC. (b) Case of (3, 1)-QRAC.

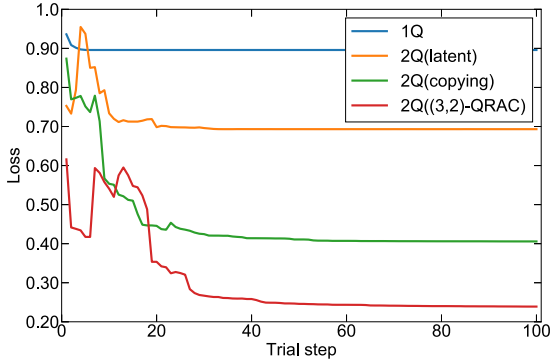

**FIG. 6.** Loss curves of four types of VQC+QRAC on type 3-4-5 over iterations. 1Q is one-qubit VQC+QRAC with (3,1)-QRAC. 2Q (latent), 2Q (copying), and 2Q ((3,2)-QRAC) represent two-qubit VQC+QRAC with latent qubits, the copies of (3,1)-QRAC, and (3,2)-QRAC, respectively.

### B. VQC WITH TWO-QUBIT QRAC

The aforementioned discussion shows that while still achieving a constant-factor saving, (3,2)-QRAC can be more powerful than the one-qubit QRAC at the expense of one more qubit used. In addition, multiqubit QRACs are therefore important in order to encode more discrete features with better efficiency. Unfortunately, a general method for constructing multiqubit QRACs beyond concatenation of one-qubit QRACs is not known, and so far only two-qubit QRACs, denoted as $(n, 2, p)$-QRACs, were studied in the literature [23], [24], [58].

Here, we investigate the practicability of VQC+QRAC using two-qubit QRACs on synthetic datasets and compare them with VQC+QRAC using (2,1)- and (3,1)-QRACs. We used $(n, 2)$-QRAC obtained in [23, Sec. III], which showed $(n, 2)$-QRAC with $n$ up to 12. To the best of our knowledge, $n = 12$ is the maximum number among all $(n, 2)$-QRACs with pure states, whose construction is explicitly known.

The synthetic datasets used in the simulation consist of $2^n$ samples of $n$ bits from $00 \ldots 0$ to $11 \ldots 1$ whose labels are determined by the following Boolean function of the bits in the two scenarios: 1-b and 2-b parity (XOR) cases. In 1-b case, the labeling is based on the $i$th bit (i.e., when the $i$th bit is 0, its label is 0, and vice versa), whereas the parity of the $i$th and $j$th bits ($i \neq j$) is used as the label in the 2-b XOR case. We should note that the indices for labeling are crucial for the classification because $(n, 2)$-QRACs are not symmetric, unlike (2,1)- and (3,1)-QRACs.

The evaluation of the VQC+QRAC on the synthetic datasets was done via the five-fold cross-validation on the $2^n$ samples for $n = 7, 8, 9, 10, 11,$ and 12. In the simulation, the indices for labeling were randomly picked. To reduce the variance of classification accuracies caused by the difference of the indices chosen for labeling, we permute the input bitstrings. More precisely, we extract a validation dataset from the training dataset, and repeatedly evaluate classification accuracies on the validation dataset with the permuted string to determine the best permutation. The train and test accuracies are calculated using the best permutation for each fold. We confirmed that this additional permutation process improves the classification accuracies. In the simulation, the number of patterns of permutation is fixed to 10. In Fig. 7(a), we see that VQC+QRAC with $(n, 2)$-QRAC performed quite well in the 1-b xor case, because in the case of $(n, 2)$-QRAC, we are guaranteed to have the POVM for recovering a single bit with probability larger than $1/2$. However, we also see that finding the measurement operator with VQC becomes more difficult as $n$ increases. From the 2-b xor case of Fig. 7(b), the VQC+QRAC could only classify correctly with probability slightly better than $1/2$ unlike the 1-b xor case.

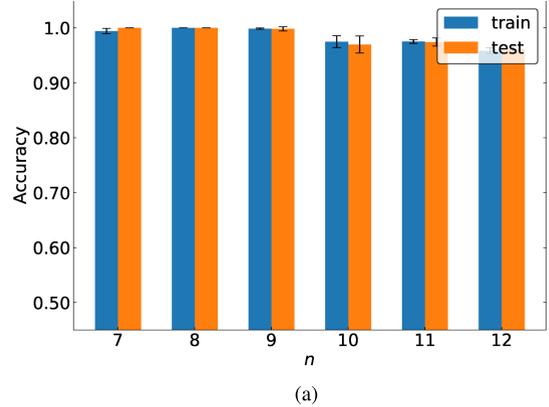

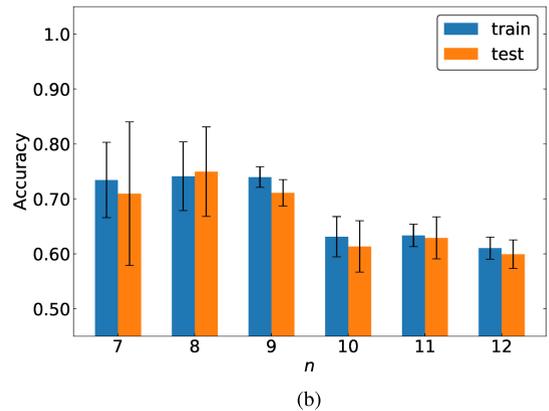

**FIG. 7.** Classification accuracies of VQC+QRAC with different $(n, 2)$-QRACs. The blue bar represents the train accuracy and the orange bar represents the test accuracy. All the values are calculated over five-fold cross-validation. Each bar and its error bar indicate the mean and standard deviation, respectively. (a) 1-b xor. (b) 2-b xor.

### C. STRATEGY WITH *f*-QRAC

Here, we show how to use the idea of *f*-QRAC in VQC+QRAC and see the effect of the encoding by solving the 2-b xor case studied in the previous section.

As reviewed in Section IV-C, the core idea of constructing *f*-QRAC is to randomly divide $n$ bits into $m$ sets and encode them with $m$ sets of $(n/m, 1)$-QRACs. Following this idea,

|  | (a) | (b) | (c) |
|---|---|---|---|
| **train accuracy** | $0.980 \pm 0.039$ | $0.571 \pm 0.014$ | $0.741 \pm 0.062$ |
| **test accuracy** | $0.977 \pm 0.046$ | $0.234 \pm 0.058$ | $0.750 \pm 0.081$ |

*Note:* (a) Four-qubit VQC with four sets of (2,1)-QRAC (depth=5-, 2-b of 2-b xor separated).
(b) Four-qubit VQC with four sets of (2,1)-QRAC [depth=5-, 2-b of 2-b xor encoded with the same (2,1)-QRAC].
(c) Two-qubit VQC with (8,2)-QRAC (depth=3).
All the values are calculated over five trials and the mean and standard deviations are shown.

we present a new method for overcoming the difficult problems, such as the 2-b xor case in Section V-B. To evaluate a Boolean function $f$ whose output depend on $k$ bits out of $n$ bits, we prepare $m$ sets of $(n/m, 1)$-QRAC for encoding the $n$ bits into $m$ qubits. Note that the length of the input bitstring of the Boolean function $k = o(\sqrt{m})$ holds to ensure that all the $k$ bits will be encoded independently with high probability. Also, we consider the cases of $n/m = 2$ or 3. Then, we apply the variational circuit to the separable output state of $m$ sets of $(n/m, 1)$-QRAC so that we can decode the value of $f$ from the resulting $m$-qubit system, whereas in $f$-QRAC, we measure each of the encoded $m$ qubits to get the substring and compute the value of $f$ classically.

Here, we conduct the same simulation of the 2-b xor case in Section V-B for $n=8$, with the four-qubit VQC+QRAC composed of four sets of (2,1)-QRAC. Note that we need more depths to optimize the VQC because more qubits are used than in the case of $(n, 2)$-QRAC. For this specific QRAC, the evaluation was performed without the permutation process because we only use the symmetric (2,1)-QRAC. The results are shown in Table I. If the two determining bits of the inputs are separated when encoding with four (2,1)-QRACs [see (a) in Table I], the classification accuracy is better than (8,2)-QRAC [see (c) in Table I]. This happens with probability 75% when $m = 4$ and $k = 2$. However, if not [see (b) in Table I], the classification accuracy is very bad as expected from the analysis in [26]. Alternatively, for example, we can use one (5,2)-QRAC and one (3,1)-QRAC to encode 8 b to reduce the number of qubits at the expense of accuracies. In this way, the concatenation of $(n, 2)$-QRACs may be useful in practice to build a $(n, m)$-QRAC for VQC.

## VI. EXPERIMENTS
We perform experiments on real-world datasets to show the possibility of QRACs for the efficient encoding of discrete features. Namely, QRACs can encode discrete features with less number of qubits and, hence, the resulting VQCs have less variational parameters. In particular, we focus on using QRACs, which encode $n$ bits into one qubit. We emphasize that no other methods exist to encode bitstrings other than the trivial basis encoding that requires more qubits. Nevertheless, here we use the nonlinear embedding in Section III-C, which here we call the ZZ feature map, to encode bitstrings for comparison with QRAC. For the experiments, we use four real-world datasets: the breast cancer (BC) dataset[1] and the heart disease (HD) dataset[2] from the UCI Machine Learning Repository [69], as well as the Titanic dataset,[3] and the MNIST dataset.[4] We note that Thumwanit *et al.* [70] also proposed a QRAC-based encoding with trainable quantum circuits and demonstrated the experiments on the Titanic and MNIST datasets. Here we use the same datasets and compare the different encoding methods.

The BC and MNIST datasets contain only discrete features and, thus, they are suitably used for comparing the feature map using QRACs to that without using QRACs. The HD and Titanic datasets contain discrete and continuous features and, thus, they can be encoded using both QRACs and the ZZ feature map.

All experiments run with simulators and an experiment on the HD dataset with a real-device backend. The real-valued (continuous) features were mapped using the ZZ feature map. Unless stated otherwise, we evaluated accuracies and training loss with the five-fold cross-validations. For all the datasets except for MNIST, the variational form we used was again the RyRz ansatz, with the depth $l = 4$ in the case of the simulator and $l = 1$ in the case of the real device to reduce the influence of the decoherence. For the MNIST dataset, we used the variational form from the tutorial in TensorFlow Quantum.[5]

The resultant train and test accuracies on all the datasets are summarized in Table II. In this article, we refer to VQC as VQC+QRAC if QRAC is used for encoding in this table. We will explain the details of each experiment in the following sections.

### A. VQC ON THE BC DATASET
The BC dataset consists of 286 instances, each of which has 9 features to predict no-recurrence or recurrence events. We removed 9 instances with missing features to obtain 196 instances of no-recurrence events, and 81 instances of recurrence events. Out of 9 features, we selected 4 features to be used in the VQC; they are *menopause* with 3 categories, *tumor-size* with 12 categories, *node-caps* with 2 categories, and *deg-malig* with 3 categories. When applying the ordinal encoding method, the VQC needs 4 qubits to map the selected features of the instances. Now note that the total number of bits of the ordinal encoding of all 4 features is 9; 2 b for menopause, 1 b for node-caps, 4 b for tumor-size, and 2 b for deg-malig. This 9-b information can then be encoded by three (3,1)-QRACs. Thus, the VQC+QRAC needs 3 qubits to map the features of the instances.

---

[1] https://archive.ics.uci.edu/ml/datasets/breast+cancer (Last accessed on May 22, 2020)
[2] https://archive.ics.uci.edu/ml/datasets/Heart+Disease (Last accessed on May 22, 2020)
[3] https://www.openml.org/d/40945 (Last accessed on May 10, 2021)
[4] http://yann.lecun.com/exdb/mnist/ (Last accessed on May 10, 2021)
[5] https://www.tensorflow.org/quantum/tutorials/mnist (Last accessed on May 10, 2021)

**TABLE II** Classification Performance of Different Encoding Methods for Several Datasets

| Dataset | Encoding | #qubits | Train accuracy | Test accuracy | Encoding detail |
|---|---|---|---|---|---|
| BC(VI-A) | ZZ | 4 | $0.698 \pm 0.021$ | $0.661 \pm 0.045$ | *menopause*, *node-caps*, *tumor-size*, and *deg-malig* |
| | (3, 1)-QRAC | 3 | $\mathbf{0.736 \pm 0.015}$ | $\mathbf{0.726 \pm 0.048}$ | *menopause*, *node-caps*, *tumor-size*, and *deg-malig* with three (3, 1)-QRACs |
| HD(VI-B1) | ZZ | 3 | $0.842 \pm 0.026$ | $0.825 \pm 0.052$ | *cp(0)*, *ca(0)*, and *thal(2)* |
| | (3, 1)-QRAC | 2 | $\mathbf{0.851 \pm 0.010}$ | $\mathbf{0.851 \pm 0.011}$ | *cp(0)*, *ca(0)*, and *thal(2)* with a (3, 1)-QRAC twice |
| HD(VI-B2) | ZZ | 4 | $0.731 \pm 0.017$ | $0.706 \pm 0.045$ | *cp(0)*, *ca(0)*, *thal(2)*, and *oldpeak* |
| | (3, 1)-QRAC+ZZ | 2 | $\mathbf{0.833 \pm 0.008}$ | $\mathbf{0.845 \pm 0.039}$ | *cp(0)*, *ca(0)*, and *thal(2)* with a (3, 1)-QRAC and *oldpeak* with ZZ |
| HD(VI-B3) | (3, 1)-QRAC+ZZ (simulator) | 2 | $0.809 \pm 0.052$ | $0.822 \pm 0.048$ | same as (3, 1)-QRAC+ZZ in HD(VI-B2) |
| | (3, 1)-QRAC+ZZ ('ibmq_almaden') | 2 | $\mathbf{0.842 \pm 0.007}$ | $\mathbf{0.861 \pm 0.035}$ | same as (3, 1)-QRAC+ZZ in HD(VI-B2) |
| Titanic(VI-C) | ZZ | 4 | $0.750 \pm 0.010$ | $0.722 \pm 0.006$ | *age*(cont.), *fare*(cont.), *sex*, and *pclass* |
| | ZZ(all dis.) | 4 | $0.736 \pm 0.008$ | $0.707 \pm 0.044$ | *age*(dis.), *fare*(dis.), *sex*, and *pclass* |
| | (3, 1)-QRAC | 3 | $\mathbf{0.770 \pm 0.023}$ | $\mathbf{0.773 \pm 0.034}$ | an ordinal encoding of *age*(dis.), *fare*(dis.), *sex*, and *pclass* with three (3, 1)-QRACs |
| | (3, 1)-QRAC+ZZ | 3 | $0.727 \pm 0.014$ | $0.717 \pm 0.028$ | an ordinal encoding of *sex* and *pclass* with a (3, 1)-QRAC and *age*(cont.) and *fare*(cont.) with ZZ |
| MNIST(VI-D) | basis encoding | 16 | $\mathbf{0.910 \pm 0.001}$ | $\mathbf{0.915 \pm 0.002}$ | a qubit for a pixel of a $4 \times 4$ image |
| | (2, 1)-QRAC | 8 | $0.836 \pm 0.005$ | $0.837 \pm 0.003$ | 16 bits of the flatten $4 \times 4$ image with eight (2, 1)-QRACs |
| | (3, 1)-QRAC | 6 | $0.880 \pm 0.005$ | $0.886 \pm 0.018$ | 16 bits of the flatten $4 \times 4$ image with six (3, 1)-QRACs |

*Note*: The classification accuracies are obtained with the five-fold cross-validation for BC, HD, and Titanic or five repetitions with different seeds for MNIST. The bold indicates the best performing value in the dataset.

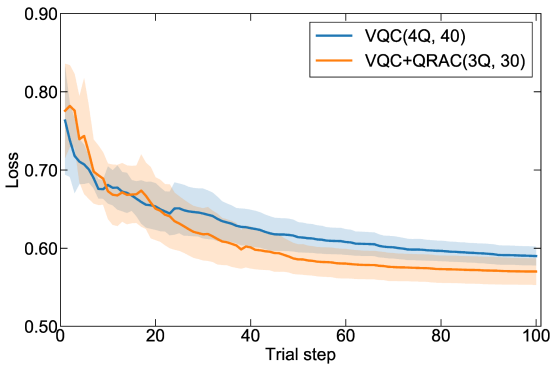

**FIG. 8.** Training loss of VQC with 4 qubits and 40 parameters (blue) and VQC+QRAC with 3 qubits and 30 parameters (orange), on the BC dataset. The solid line is the mean of the loss values, and the shading represents their standard deviation over five-fold cross-validation.

We compared the VQC and the VQC+QRAC by the five-fold cross-validation on the BC dataset. Notice that the target labels are unbalanced. For this reason, we trained both models with oversampling to balance the training dataset, but we applied the model on the unbalanced data to evaluate the test accuracy. Fig. 8 shows the average and standard deviation of the training loss of both VQC and VQC+QRAC. We can see from the figure that the VQC+QRAC achieved lower training losses. The classification accuracy of the VQC+QRAC is $0.736 \pm 0.015$ for training, and $0.726 \pm 0.048$ for testing, as shown in Table II. The accuracies are about the same as those obtained by the previous work [71]. In addition, the F1 score is a more accurate metric for the evaluation on unbalanced datasets. The F1 score on the test data is $0.216 \pm 0.086$ and $0.527 \pm 0.095$ for the case of the VQC and the VQC+QRAC, respectively.

### B. VQC ON THE HD DATASET

The HD dataset consists of 303 instances, each of which has 13 features to predict the presence of heart disease in the patient. It includes both continuous and discrete features. We tested the effectiveness of VQC+QRAC in two scenarios. First, we performed a feature engineering to discretize the real-valued features, which is often done when using tree-based classifiers. After the feature engineering, the classification task by VQC+QRAC can be conducted, similarly as in the case of BC dataset. The second is when we combine discrete and continuous features, i.e., when we use QRAC in conjunction with the ZZ feature map. We represent this encoding as QRAC+ZZ in Table II.

#### 1) CLASSIFICATION WITH BINARY FEATURES

For the first scenario, we turned real-valued features, such as *oldpeak*, into binary features by partitioning them with their median values, and applied the one-hot encoding method for discrete features, such as *cp* and *thal*. We then took the three most important features based on their importance estimated by a random forest classifier. The selected three features are chest pain type *cp(0)*, number of major vessels colored by fluoroscopy *ca(0)*, and thallium heart scan *thal(2)*. Here, for example, the discrete feature cp was transformed into the binary feature cp(0), which is true if cp is equal to 0 and false if not. The VQC used three qubits for mapping the aforementioned features, whereas the VQC+QRAC can encode them with a single (3, 1)-QRAC. Nevertheless, we used two (3,1)-QRACs for VQC+QRAC to encode the same 3-b features in order to increase the dimension of the Hilbert space. Thus, the VQC+QRAC used two qubits for mapping the aforementioned features. This is the technique proposed in Section V-A.

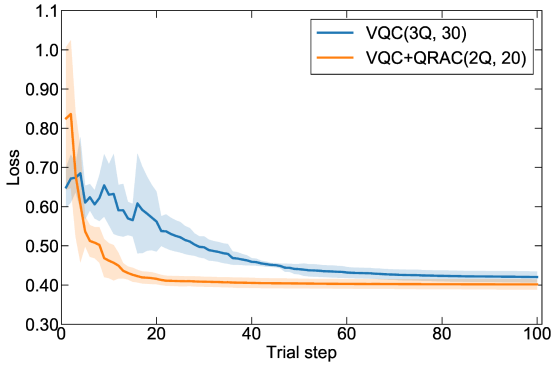

**FIG. 9.** Training loss of VQC with 3 qubits and 30 parameters (blue) and VQC+QRAC with 2 qubits and 20 parameters (orange), on the HD dataset. The solid line is the mean of the loss values, and the shading represents their standard deviation over five-fold cross-validation.

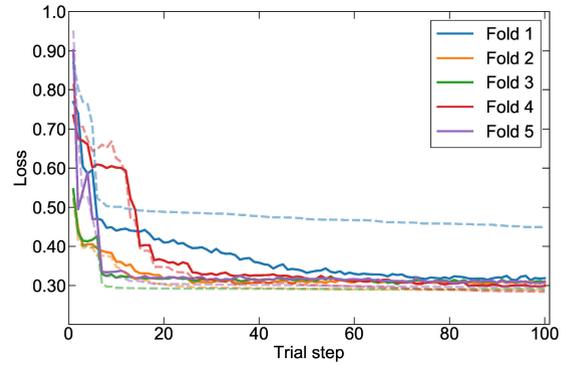

**FIG. 10.** Training loss of VQC+QRAC with the real device "ibmq_almaden" (solid) and the simulator (dashed), on the HD dataset. Each color represents one fold out of five-fold cross-validation.

The experimental results are shown in Fig. 9 and Table II. We can confirm that VQC+QRAC achieved slightly better accuracies, and its training loss saturated faster than VQC, although it started with the higher value. Less number of qubits and parameters in the variational circuit of VQC+QRAC is one of the reasons for its faster convergence. Furthermore, its higher accuracies hint that QRAC can map discrete features onto the Hilbert space in a geometrically good manner, despite the fact that the dimension of Hilbert space of the VQC+QRAC is less than VQC.

### 2) CLASSIFICATION WITH BINARY AND REAL-VALUED FEATURES

For the second scenario, we used the previous 3-b binary features in addition to the real-valued feature oldpeak, which represents ST depression induced by exercise relative to rest. The VQC used four qubits, whereas the VQC+QRAC used two qubits; one qubit to encode the 3-b binary features with a single (3,1)-QRAC and the other for oldpeak.

From Table II, we can see that the addition of the real-valued features did not increase the accuracies of the VQC. In fact, both VQCs had lower accuracies with the addition of the new feature, possibly because of the more number of parameters in the variational circuits. However, we can see that the VQC+QRAC still performed well and its accuracies are comparable with the case using only binary features.

### 3) CLASSIFICATION ON QUANTUM DEVICES

Having demonstrated that VQC+QRAC can achieve better results to map discrete and continuous features on simulators, we then performed the experiment on a quantum processor

"ibmq_almaden" using its first and second physical qubits. The experiment was executed from May 10 until May 19, 2020 through fair-share queuing policy of the IBM Quantum Systems.

The experimental results are shown in Fig. 10 and Table II. The depth $l = 1$ was used in the experiment, and the number of shots was 1024 for classifying one instance. We can see that the quantum processor achieved almost the same performance as the simulator. The reason the quantum processor performed slightly better than the simulator can be explained by Fig. 10; that is, whereas the optimization of Fold 1 on the simulator (blue dashed line) seems to be stacked at local minima, the optimizer with the quantum processor (blue solid line) circumvents being trapped in such a plateau. This might be caused by the noise in the real device, but except for Fold 1, we confirmed that the optimization with the quantum processor was done very similarly to the simulator.

### C. VQC ON THE TITANIC DATASET

The Titanic dataset with labels consists of 891 instances, each of which has 11 features, and the target variable is the survival of the passengers. The four most important features for this experiment are selected through a random forest classifier. They are mixture of discrete features (*sex* and *pclass*) and continuous ones (*age* and *fare*).

We compared four different encodings, as shown in Table II. These include cases where the continuous features are discretized, represented as (dis.) in the table, and cases where they are not, represented as (cont.). The continuous features are rescaled and the ordinal encoding method is applied, as done in the BC dataset, to the discrete features. From Table II, we can see that VQC+QRAC without the ZZ feature map obtains the best accuracies.

### D. VQC ON THE MNIST DATASET

MNIST is a dataset of handwritten digit images of size $28 \times 28$. For binary classification, only images for "3" and "6" were selected, whereas those labeled both "3" and "6" were omitted. The resulting number of instances is 11 520 for training and 1968 for testing. For our experiment, the selected images were downsampled to $4 \times 4$ by performing bilinear interpolation and the pixel values are binarized. We used the variational form consisting of Ising $XX$ gates and Ising $ZZ$ gates, with Adam as an optimizer. We followed the aforementioned preprocessing and the variational form from the tutorial in TensorFlow Quantum. We compared three encodings, as shown in Table II.

Table II shows the average accuracies on five different runs. We can see that although the accuracies of the VQC+QRACs are slightly worse than the VQC with basis encoding, they require less number of qubits. A technique of using QRAC for the convolutional embedding to improve accuracies is shown in [70].

### E. REMARKS ON TRAINABILITY OF VQC+QRAC

Here, we provide a general discussion on the trainability of quantum machine learning scheme. It is known that variational quantum algorithms, including VQC, can exhibit barren plateaus, where the cost gradient will vanish exponentially with the number of qubits [72]. This phenomenon causes the difficulty of training variational circuits. Hence, there have been several approaches proposed that explore an ansatz of parameterized circuits or the form of optimizer, to circumvent the barren plateau issue. A solution might be limiting the circuit depth; in particular, Cerezo et al. [73] proposed to use a special type of shallow circuit called the alternating layered ansatz, which was later proven to have (enough) expressibility [74].

In this perspective, our approach can be regarded as the method that may circumvent the barren plateau issue by limiting the circuit width (i.e., the number of qubits). The point of our method is that the QRAC theory gives a guide for directly decreasing the number of qubits used for encoding discrete dataset. Obviously, too much limitation will decrease the expressibility of the circuit and consequently the classification performance, as demonstrated by the case of $(n, 1)$-QRAC discussed in Section V-A. To keep both the trainability and the expressibility of the circuit, using the $(n, m)$-QRAC or $(n, m)$-$f$-QRAC with, e.g., $m = n/2$ will work, as indicated by the demonstration given in Sections V-B and V-C. Note that the size of the example problems considered in this article is very small, and thus the barren plateau issue is not clearly observed. We will study large-size problems to prove the genuine advantage of our QRAC-based method as a variational quantum circuit having both the trainability and the expressibility.

### VII. CONCLUSION

This article proposed the QRAC-based quantum classifier for a given dataset having discrete features. The main advantage of the scheme is to provide the mean for encoding an input bitstring to a quantum state with less number of quantum bits: more precisely, the QRAC theory guarantees that we can encode a bitstring of length $n$ into $\log(n)/2$ qubits and recover any one out of $n$ bits with probability bigger than $1/2$. This results in a shorter circuit for learning the classification task by decreasing the number of parameters in variational circuits. This is advantageous in the current status of this research field where only relatively small-size quantum computers are available. After our work was published, there have been other evidences of the high efficiency of QRAC-based quantum classifiers on other datasets [70], [75]. Those demonstrate the generality of our proposal. Also, even when an ideal fault-tolerant quantum computer emerges, the proposed scheme still has a clear merit in that, compared to quantum classifiers without QRAC, a shorter circuit is easier to train, which hopefully may realize better classification performance. This practical advantage was in fact observed in the numerical simulation demonstrated in this article.

As described in Section IV-A, originally, QRAC is a theory providing a solid quantum advantage over the classical one, in the problem for probabilistically extracting (1 b) information by appropriately synthesizing quantum measurement. Although in this article, we only utilize the encoding part of this QRAC theory, we will further combine the probabilistic information-extraction aspect of QRAC to extend the proposed method so that it could have a certain quantum advantage in the machine learning context. Additionally, analyzing the robustness of the proposed technique itself and its tolerance to the noise is left for future work.

### ACKNOWLEDGMENT

The authors thank N. Thumwanit of The University of Tokyo for his assistance on the experiments. The authors also thank the reviewers for their sincere feedback to improve the quality of this article.

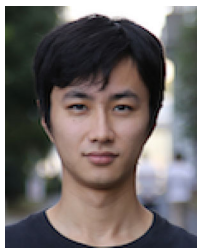

**Hiroshi Yano** is currently working toward the Ph.D. degree in applied physics and physicoinformatics with Keio University, Yokohama, Japan.

His research interests include quantum machine learning and noisy intermediate-scale quantum algorithms.

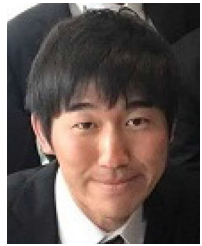

**Yudai Suzuki** is currently working toward the Ph.D. degree in mechanical engineering with the School of Science for Open and Environmental Systems, Keio University, Yokohama, Japan.

His research interests include quantum machine learning and noisy intermediate-scale quantum algorithms, particularly quantum kernel method, variational quantum algorithms, and quantum reservoir computing.

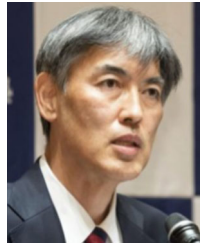

**Kohei M. Itoh** received the B.S. degree in instrumentation engineering from Keio University, Yokohama, Japan, and the M.S. and Ph.D. degrees in materials science and mineral engineering from the University of California, Berkeley, Berkeley, CA, USA, in 1989, 1992, and 1994, respectively.

He joined Keio University as a Faculty Member in 1995 and became a Full Professor in 2007. He served as the Dean of Faculty and Graduate School of Science and Technology, Keio University between 2017 and 2019, and as the Chair of Keio AI and Advanced Programming Consortium between 2018 and 2021. His main focus of research has been quantum computing and quantum sensing.

Dr. Itoh was a recipient of the Japan IBM Prize in 2006 and the Japan Society for the Promotion of Science Prize in 2009. He is also a Founder of the IBM Quantum Computer Network Hub at Keio University.

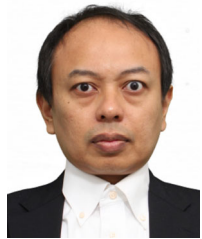

**Rudy Raymond** received the Ph.D. degree in quantum algorithms from the Graduate School of Informatics, Kyoto University, Kyoto, Japan, in 2006.

He is currently a Researcher with IBM Research—Tokyo, Tokyo, Japan. His areas of expertise include artificial intelligence, optimization, quantum information, and quantum computing.

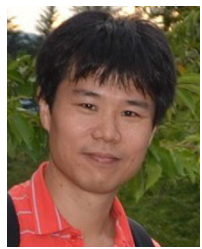

**Naoki Yamamoto** received the B.S. degree in engineering and the M.S. and Ph.D. degrees in information physics and computing from The University of Tokyo, Tokyo, Japan, in 1999, 2001, and 2004, respectively.

From 2004 to 2006, he was a Visiting Researcher with the California Institute of Technology, Pasadena, CA, USA. From 2007 to 2008, he was a Research Fellow with The Australian National University, Canberra, ACT, Australia. He is currently a Professor with Keio University, Yokohama, Japan. His research interests include quantum computation and control.